\documentclass[12pt]{iopart}
\usepackage{iopams}  
\usepackage{epsfig}
\usepackage{graphicx}

\begin{document}

\title[Initial State Granularity]{A systematic study of the sensitivity of
triangular flow to the initial state fluctuations in relativistic heavy-ion
collisions}
\author{Hannah Petersen${}^1$, Rolando La Placa${}^{1,2}$ and Steffen A.
Bass${}^1$ \\[.4cm]}

\address{${}^1$~Department of Physics, Duke University, Durham, North Carolina
27708-0305, United States\\
${}^2$~Harvard University, Cambridge, Massachusetts 02138, United States}

\ead{hp52@phy.duke.edu}

\begin{abstract}
Experimental data from the Relativistic Heavy Ion Collider
(RHIC) suggests that the quark gluon plasma behaves almost like an ideal fluid.
Due to its short lifetime, many QGP properties can only be inferred
indirectly through a comparison of the final state measurements with transport
model calculations. Among the current phenomena of interest
are the interdependencies between two collective flow phenomena, elliptic and
triangular flow. The former is mostly related to the initial geometry
and collective expansion of the system whereas the latter is sensitive to the
fluctuations of the initial state. For our investigation we use a
hybrid transport model based on the Ultra-relativistic Quantum Molecular
Dynamics (UrQMD) transport approach  using an ideal hydrodynamic expansion for
the hot and dense stage. Using UrQMD initial conditions for an Au-Au collision,
particles resulting from a collision are mapped into an energy density
distribution that is evolved event-by-event with a hydrodynamic calculation.
By averaging these distributions over different numbers of events, we have
studied how the granularity/smoothness of the distribution affects the initial
eccentricity, the initial triangularity, and the resulting flow components. The
average elliptic flow in non central collisions is not sensitive to the
granularity, while triangular flow is. The triangularity might thus provide a
good measure of the amount of initial state fluctuations that is necessary to
reproduce the experimental data.
\end{abstract}




\maketitle

\section[Introduction]{Introduction}
\label{intro}

Higher order flow coefficients have recently been recognized as a new observable to gain information about the creation of the quark gluon plasma in relativistic heavy ion reactions and its properties \cite{Agakishiev:2010ur,Alver:2010dn,Schenke:2010rr,Petersen:2010cw,Mocsy:2010um,Gardim:2011xv}. Elliptic flow, the second order Fourier coefficient of the azimuthal distribution of the particles in the final state, has been investigated in detail during the last decade, since  it is the crucial observable that is responsible for the proof that the quark gluon plasma is a nearly perfect liquid \cite{Song:2010mg}. Pressure gradients translate the initial state coordinate space eccentricity to the final state momentum space ellipticity and this connection is affected by the viscosity and the equation of state. 

Triangular flow and even higher moments of the Fourier decomposition have been measured during the last year by various experimental collaborations at RHIC and LHC \cite{Adare:2011tg,ALICE:2011ab,Chatrchyan:2011eka}. The higher odd anisotropic flow coefficients require the treatment of event-by-event fluctuations, since they vanish averaged over events. A lot of recent theoretical development has been aimed towards a better understanding of the correlations between initial and final state. Most of these models work in a simplified framework, using e.g. linearized hydrodynamics or rotated averaged initial conditions \cite{Teaney:2010vd,Qin:2010pf,Qiu:2011iv,Staig:2011wj}. 

The initial state of a heavy ion reaction that is needed to run a (viscous) hydrodynamic evolution is usually modeled following a Glauber or CGC approach \cite{Qiu:2011hf}. Also more realistic initial conditions have been investigated using event generators like NEXUS, URQMD, etc \cite{Bleicher:1998wd,Grassi:2005pm}. Some people have also studied simplified models like the one-tube model \cite{Andrade:2010xy} or just a statistical number of hot spots to get a more systematic handle on understanding initial state structures \cite{Bhalerao:2011bp,Qin:2011uw}. 

Between all these different approaches there seems to be agreement, that the flow coefficients can be used to constrain the initial state dynamics and have the potential to sort out different models \cite{Lacey:2011ug,Schenke:2011bn}. The ultimate goal is to understand the initial energy deposition which has fundamentally to do with the distributions of the nucleons/partons in the incoming nuclei and the interactions they are undergoing.  

The systematic study that is presented in this paper constitutes one step on the way to constrain the initial state granularity. By using a hybrid transport approach that is based on the Ultra-relativistic Quantum Molecular Dynamics including an (3+1) dimensional ideal hydrodynamic expansion, we demonstrate that triangular flow is directly related to the amount of fluctuations in the initial state. By using a full event-by-event setup we can calculate the triangular flow via the event plane method as it is measured by the PHENIX collaboration and put some first constraints on the initial state granularity. 

In Section \ref{hybrid} the hybrid model is described and it is explained how different granularities are obtained by initially averaging over different numbers of events. The following Section \ref{basics} demonstrates that the chosen setup does not change the bulk properties of the system by looking at pion and kaon transverse mass spectra and elliptic flow in non-central collisions. After that, we come back (in Section \ref{en_dist}) to the initial state eccentricity and triangularity distributions as a function of granularity leading directly to the resulting anisotropic flow coefficients in Section \ref{vn_data}. In the end, we will draw conclusions from the presented results and outline further research.    

\section[Model Description]{Initial State Granularity}
\label{hybrid}

The initial state profile and its granularity are mainly influenced by the following three
things: the shape of the incoming nuclei, the interaction mechanism and the distribution
of binary collisions and the degree of thermalization. In our approach all of these are given by a dynamic transport model, namely Ultra-relativistic Quantum Molecular Dynamics (UrQMD) \cite{Bass:1998ca,Bleicher:1999xi,Petersen:2008dd}. The incoming nucleons are sampled according to Woods-Saxon profiles and the initial nucleon-nucleon scatterings and non-equilibrium dynamics proceed
according to the Boltzmann equation. After the two nuclei have passed through
each other, at  $t_{\rm start}=0.5$ fm, local thermal equilibrium is assumed to perform the transition to the ideal hydrodynamic
description.

The particle degrees of freedom are represented by three-dimensional
Gaussian distributions in the following way
\begin{equation}
\epsilon(x,y,z)=\left(\frac{1}{2\pi}\right)^{\frac{3}{2}}\frac{\gamma_z}{
\sigma^3} E_p \exp{-\frac{(x-x_p)^2+(y-y_p)^2+(\gamma_z(z-z_p))^2}{2\sigma^2}}
\end{equation}
to obtain energy, momentum and net baryon density distributions that are smooth
enough for the hydrodynamic evolution. Here, $\epsilon$ is the energy density at
position $(x,y,z)$ that a particle with energy $E_p$ at position $(x_p,y_p,z_p)$
contributes. The Gaussians are Lorentz contracted in z-direction by $\gamma_z$
to account for the large longitudinal velocities. Only the matter at midrapidity
$|y|<2$ is assumed to be locally equilibrated, whereas the other hadrons are
treated in the hadronic cascade. 

A similar procedure is often employed in Monte Carlo approaches for hydrodynamic initial conditions. The width of the Gaussians representing the individual particles or tubes that lead to hot spots in the initial state is then varied and the granularity changes. The disadvantage of adjusting this parameter is that the overall features of the initial state change as well, e.g. the entropy density or the maximum value of the profile and need to be re-adjusted. In \cite{Petersen:2010zt} we have shown how bulk observables are affected by variations of the obvious initial state parameters, the starting time and the Gaussian width. 

For the present systematic study we have chosen a different path: By averaging over different numbers ($n=1$ - $n=25$) of initial events at the same beam energy, impact parameter and collision system, the bulk properties of the system are kept constant, while the granularity is varied. The disadvantage of the averaging procedure is that the direct connection to the initial state physics is lost. Still, this setup is a powerful tool to show systematic dependencies on the amount of initial state fluctuations. 

\begin{figure}[ht]
\hspace{-3cm}
\begin{minipage}[b]{0.3\linewidth}
\centering
\includegraphics[scale=0.7]{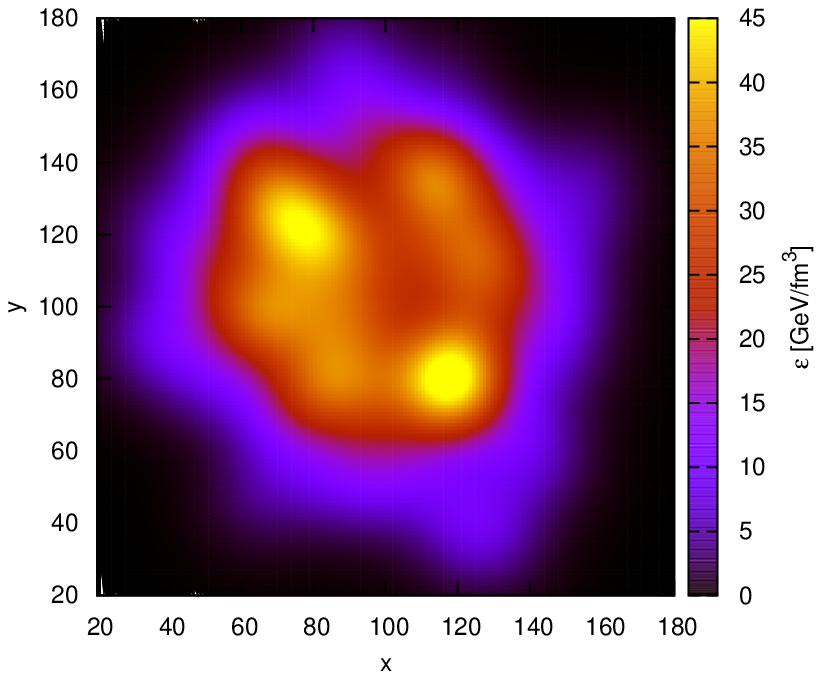}
\end{minipage}
\hspace{1cm}
\begin{minipage}[b]{0.3\linewidth}
\centering
\includegraphics[scale=0.7]{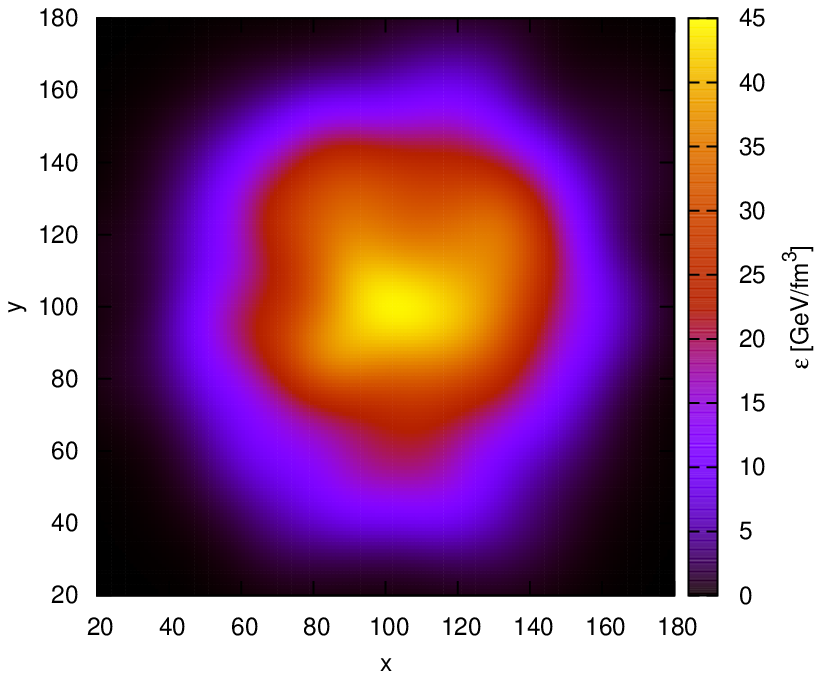}
\end{minipage}
\hspace{1cm}
\begin{minipage}[b]{0.3\linewidth}
\centering
\includegraphics[scale=0.7]{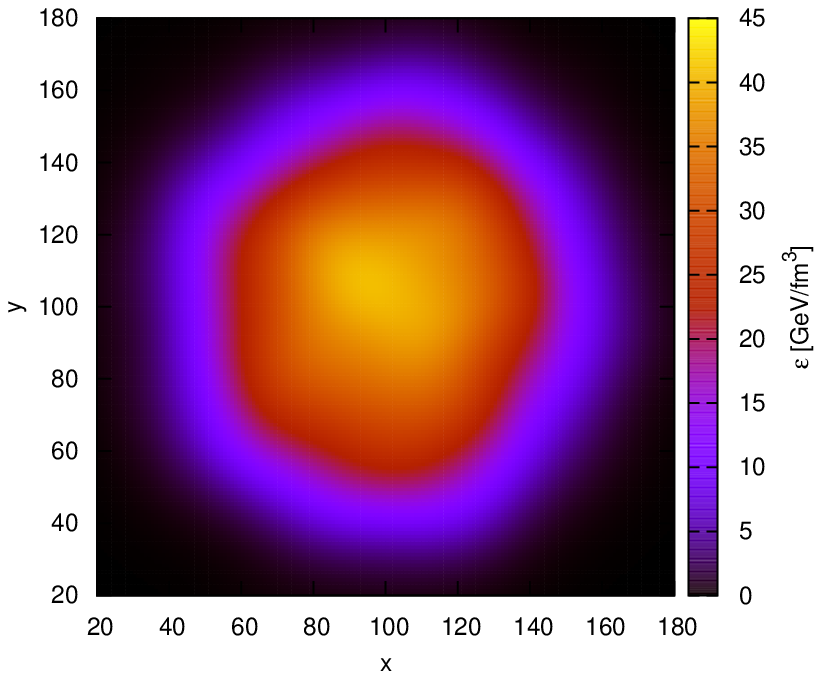}
\end{minipage}
\caption{Energy density distributions in the transverse plane for central ($b=2$ fm) Au+Au collisions at $\sqrt{s_{\rm NN}}=200$ GeV with three different granularities. The left picture depicts the profile from one event, the middle one has been averaged over 5 events and the right panel is a smooth average of 25 events.}
\label{fig_ini_edens2}
\end{figure}

\begin{figure}[ht]
\hspace{-3cm}
\begin{minipage}[b]{0.3\linewidth}
\centering
\includegraphics[scale=0.7]{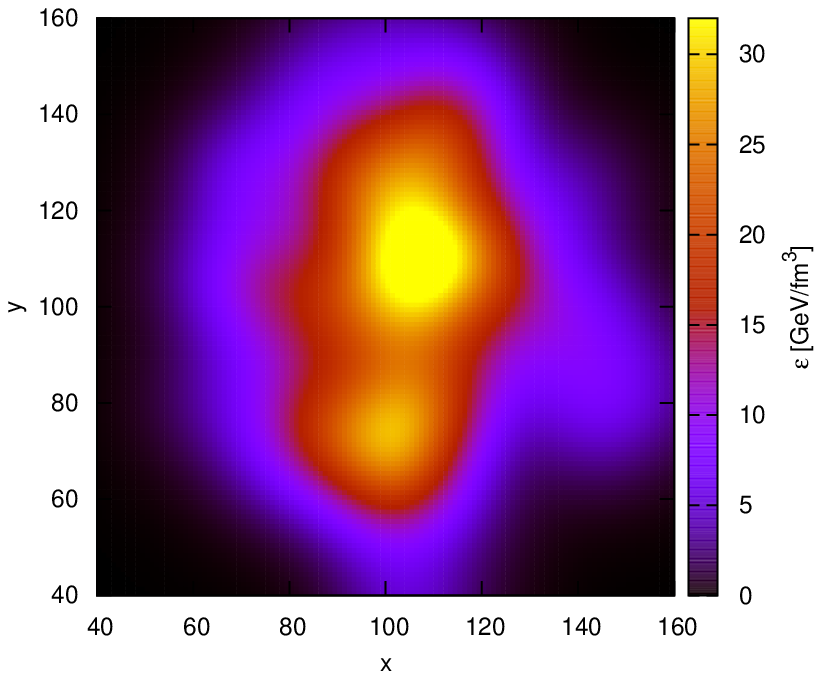}
\end{minipage}
\hspace{1cm}
\begin{minipage}[b]{0.3\linewidth}
\centering
\includegraphics[scale=0.7]{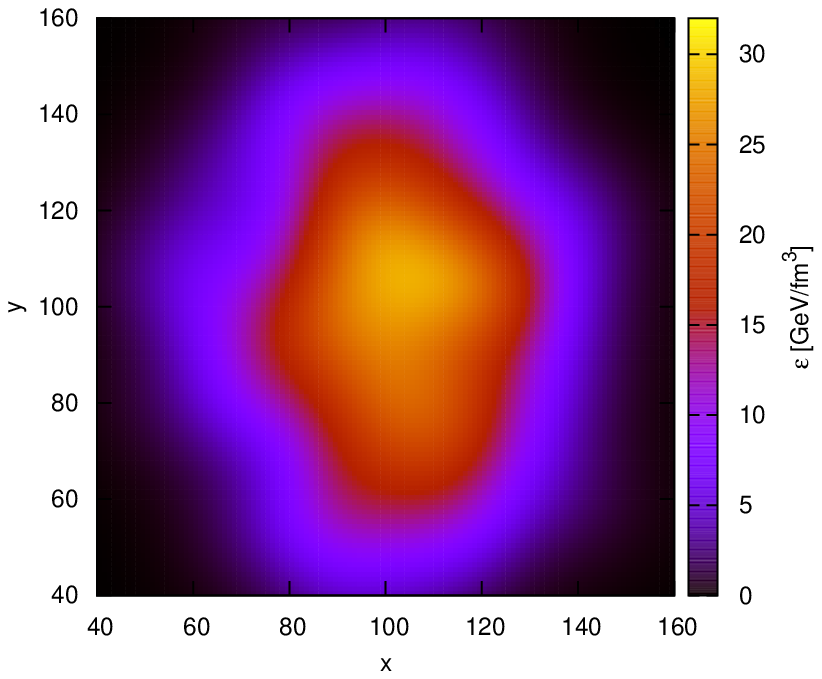}
\end{minipage}
\hspace{1cm}
\begin{minipage}[b]{0.3\linewidth}
\centering
\includegraphics[scale=0.7]{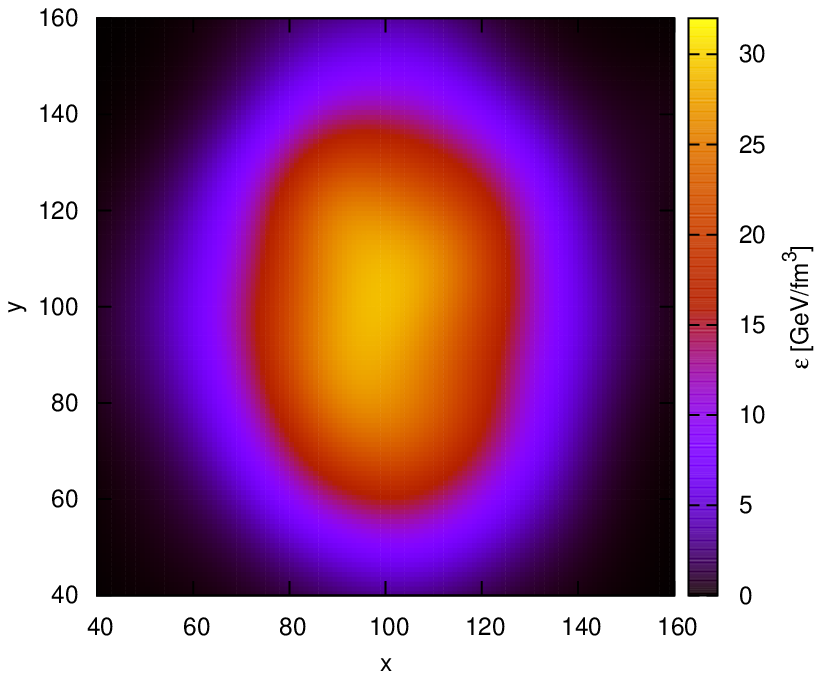}
\end{minipage}
\caption{Energy density distributions in the transverse plane for mid-central ($b=7$ fm) Au+Au collisions at $\sqrt{s_{\rm NN}}=200$ GeV with three different granularities. The left picture depicts the profile from one event, the middle one has been averaged over 5 events and the right panel is a smooth average of 25 events.}
\label{fig_ini_edens7}
\end{figure}

Fig. \ref{fig_ini_edens2} and \ref{fig_ini_edens7} show examples for initial states that are generated by averaging over 1, 5 or 25 events. The maximum energy density in the transverse plane in central and non-central collisions is similar in the three cases and the granularity differs from a maximum amount of fluctuations using the full event-by-event setup ($n=1$) to pretty smooth initial conditions ($n=25$). We have generated 300 of each of these initial profiles for 6 different granularities for Au+Au collisions at $\sqrt{s_{\rm NN}}=200A$ GeV at two impact parameters.

These individual initial conditions are propagated through a full (3+1)
dimensional ideal hydrodynamic evolution \cite{Rischke:1995ir,Rischke:1995mt} and a subsequent hadronic afterburner. For the equation of state and the freeze-out criterion the most successful values found in \cite{Petersen:2010zt} have been used (DE-EoS and $\epsilon<=1.022$ GeV/fm$^3$).

The event-by-event calculation provides the full final phase-space distribution
of the hadrons that are also measured in experiments and therefore we are able to employ the same event plane measurement method as the PHENIX collaboration has used for their $v_3$ analysis. This allows us to directly compare to experimental data and investigate the sensitivity to the granularity.  

\section[Basic Checks]{Spectra and Elliptic Flow}
\label{basics}

\begin{figure}[h]
\includegraphics[width=0.5\textwidth]{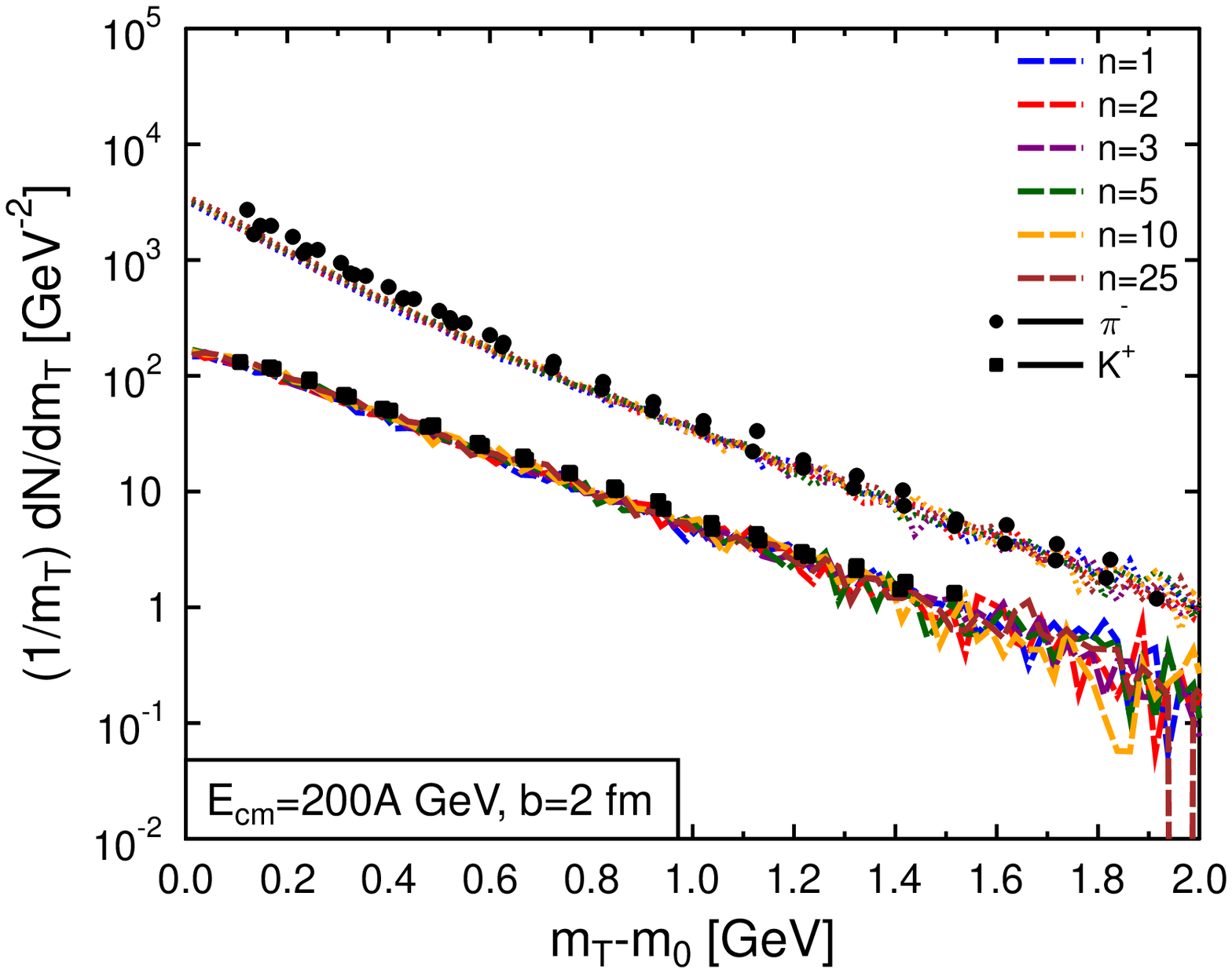}
\includegraphics[width=0.5\textwidth]{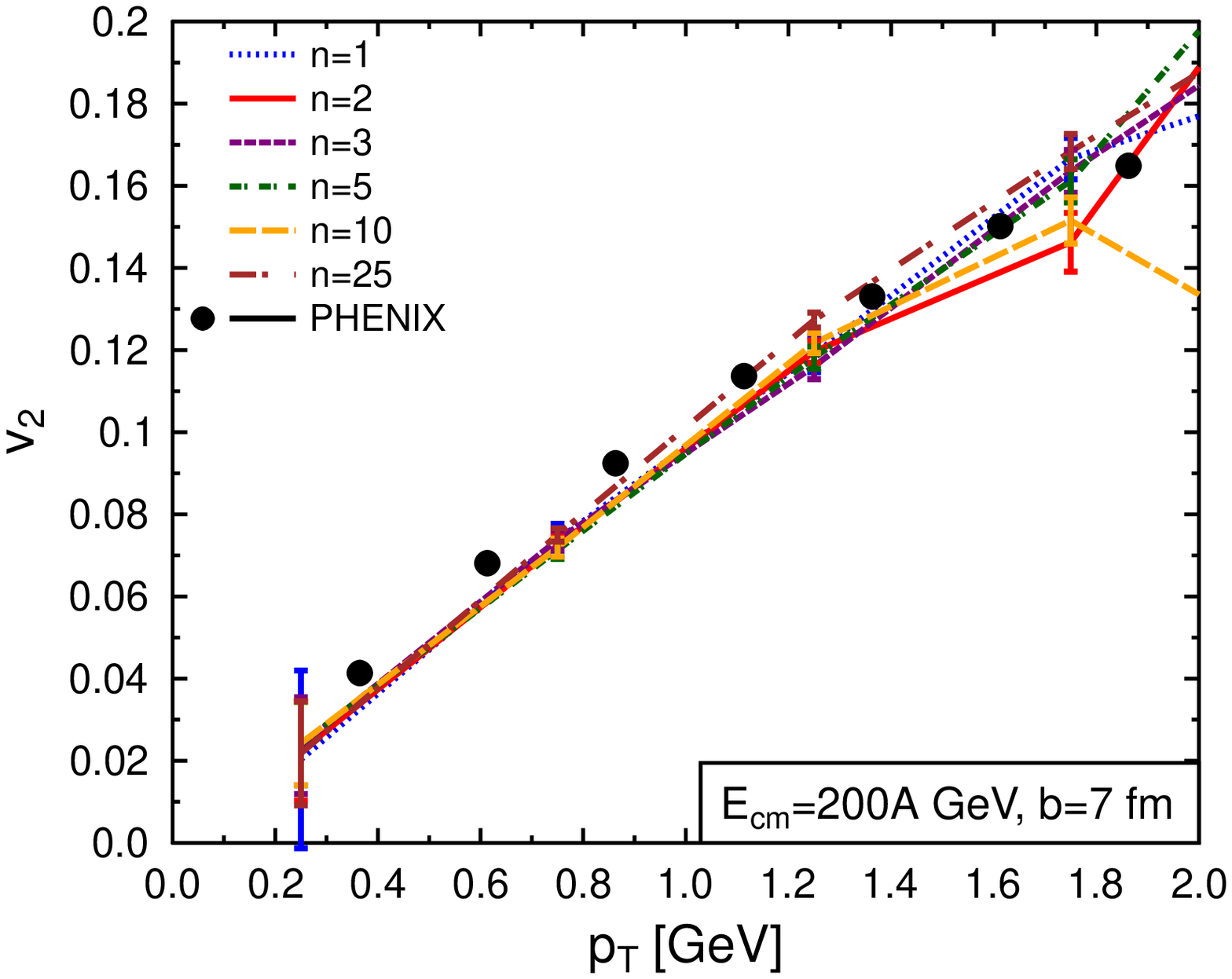}
\caption[Transverse mass spectra and elliptic flow]{Transverse mass spectra for
$\pi^-$ and $K^+$ (left) and elliptic flow as a function of transverse momentum
for charged particles (right) at midrapidity
($|y|<0.5$) in central/mid-central ($b=2$ fm/$b=7$ fm) Au+Au collisions at
$\sqrt{s_{\rm NN}}=200$ GeV from the hybrid approach for different granularities
compared to experimental data 
\cite{Adams:2003xp,Adler:2003cb,Arsene:2005mr,Adare:2011tg}. \label{fig_dndmt_flow} }
\end{figure}

In Fig. \ref{fig_dndmt_flow} a comparison between the hybrid approach for different granularities and the experimental data for basic bulk observables is shown. The left hand figure depicts transverse mass spectra for pions and kaons in central collisions and the right hand figure displays elliptic flow for charged particles as a function of transverse momentum in mid-central collisions.  

The elliptic flow has been calculated using the event-plane method including corrections to remove auto-correlations and resolution as described in \cite{Petersen:2010cw} which matches the PHENIX analysis \cite{Adare:2011tg}. The major observation is that the transverse mass spectra and the elliptic flow do not depend on the granularity. This result proves that our setup leaves the bulk properties unchanged \cite{Petersen:2010md} and is in reasonable agreement with the experimental data. 

\section[Eccentricity and Triangularity]{Eccentricity and Triangularity Distributions}
To exploit the potential sensitivity of final state flow coefficients on the granularity let us first have a look at the initial state coordinate space Fourier coefficients. The eccentricity and the triangularity have been calculated from the center transverse slice ($z=0 fm$) of the initial energy density distribution using the following formula: 
\begin{equation}
\epsilon_n=\frac{\sqrt{\langle r^n \cos(n \phi)\rangle^2+\langle r^n \sin(n
\phi)\rangle^2}}{\langle r^n \rangle} \quad \mbox{.}
\end{equation}
$\langle,\rangle$ depicts an average over the cells of the hydrodynamic grid weighted by the energy density of the respective cell and $r$ and $\phi$ are polar coordinates.

\label{en_dist}
\begin{figure}[h]
\includegraphics[width=0.5\textwidth]{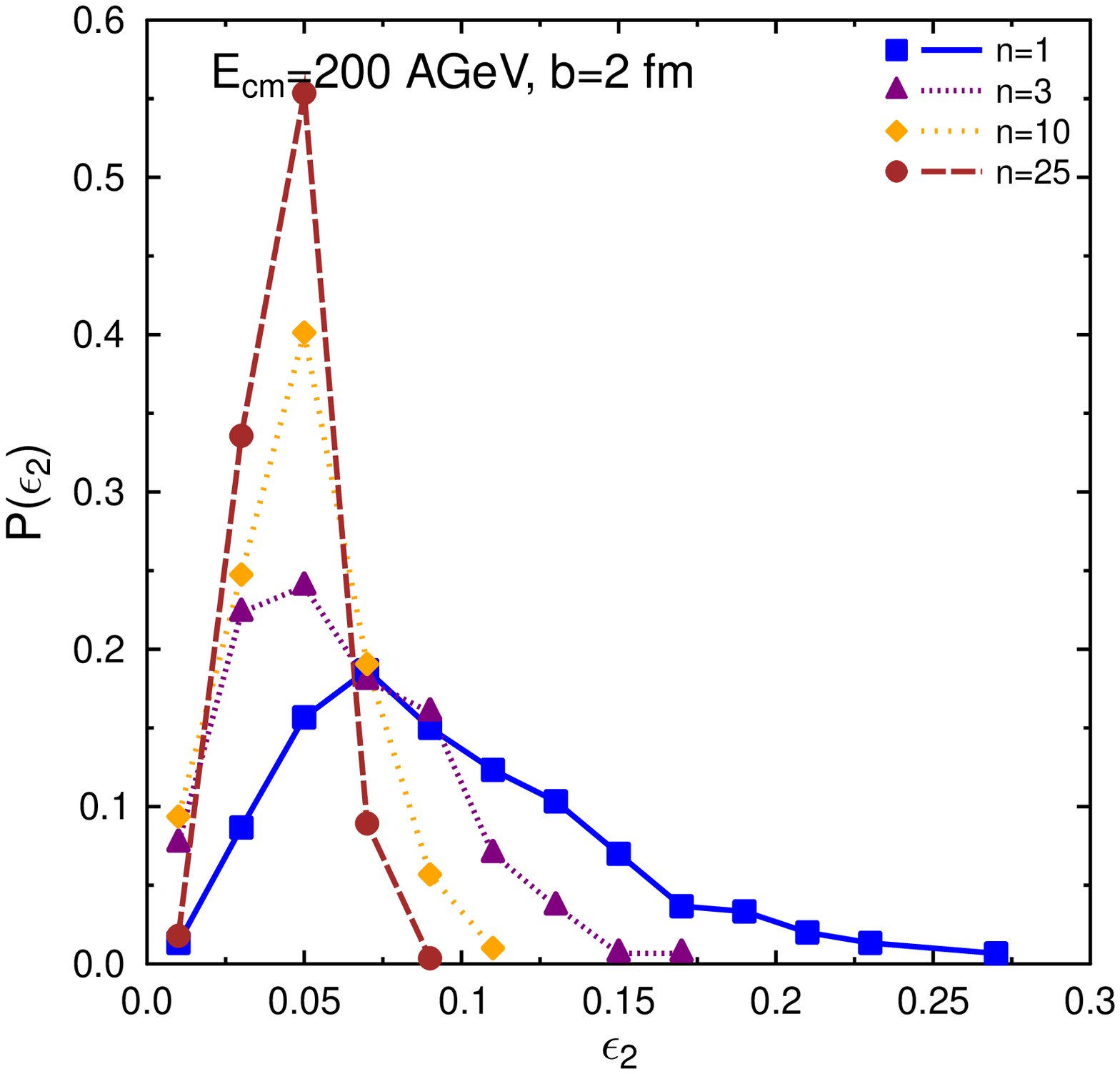}
\includegraphics[width=0.5\textwidth]{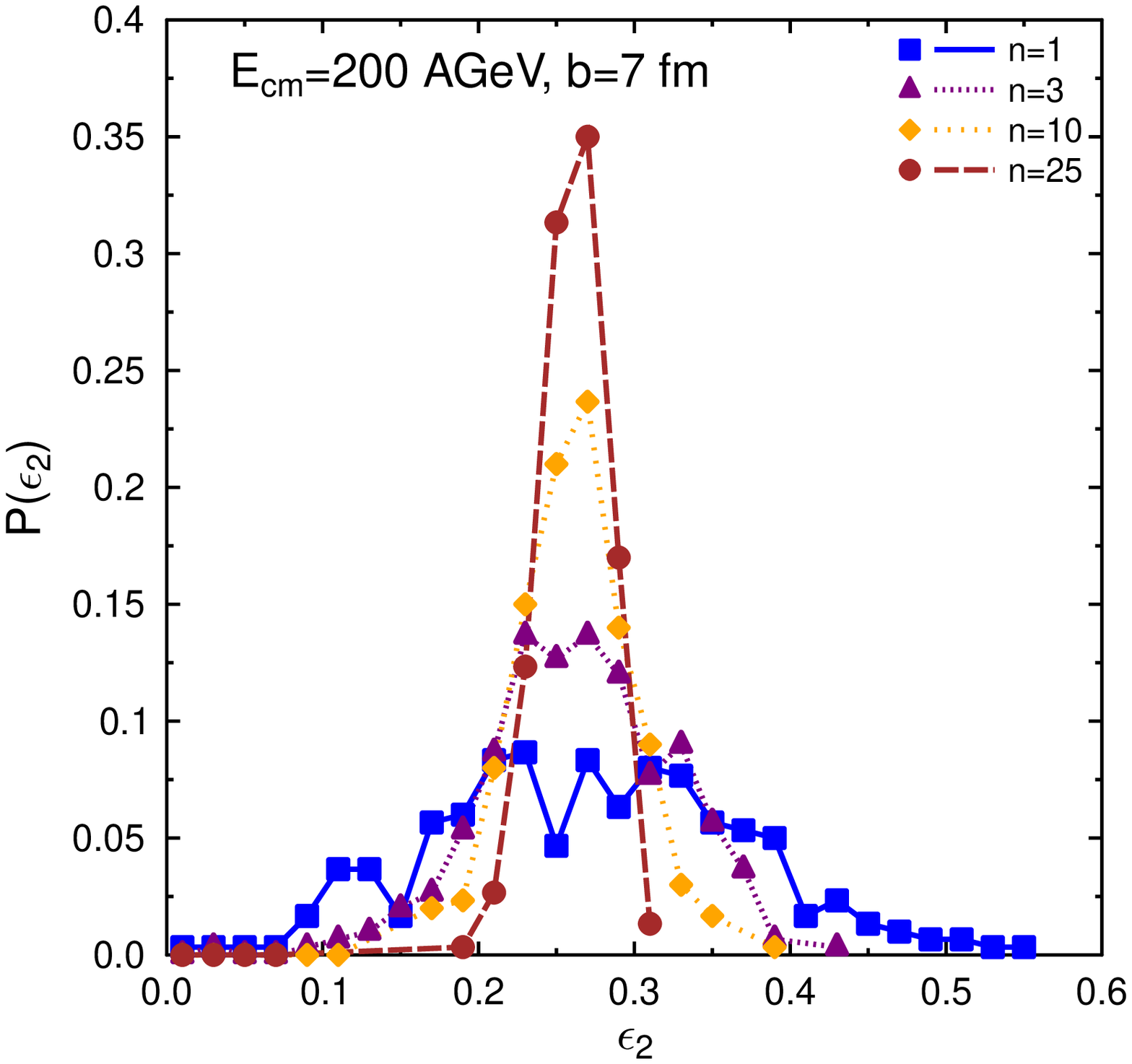}
\caption[Eccentricity distributions]{Eccentricity distributions for different
granularities in central/mid-central (left: $b=2$ fm and right: $b=7$ fm) Au+Au collisions at
$\sqrt{s_{\rm NN}}=200$ GeV.}
\label{fig_ecc_dist} 
\end{figure}

In Fig. \ref{fig_ecc_dist} the distribution of the eccentricity for 6 different granularities in central (left) and mid-central (right) collisions is shown. In the later case, the mean value of $\epsilon_2$ is remarkably stable and the distribution gets only more and more peaked for smoother initial conditions. This behavior is consistent with the constant elliptic flow result in the previous Section. For central collisions the major source of elliptic flow are initial state fluctuations and not the geometry given by the reaction plane as in non-central collisions, therefore the mean value of eccentricity decreases for smoother more central collisions.  

\begin{figure}[h]
\includegraphics[width=0.5\textwidth]{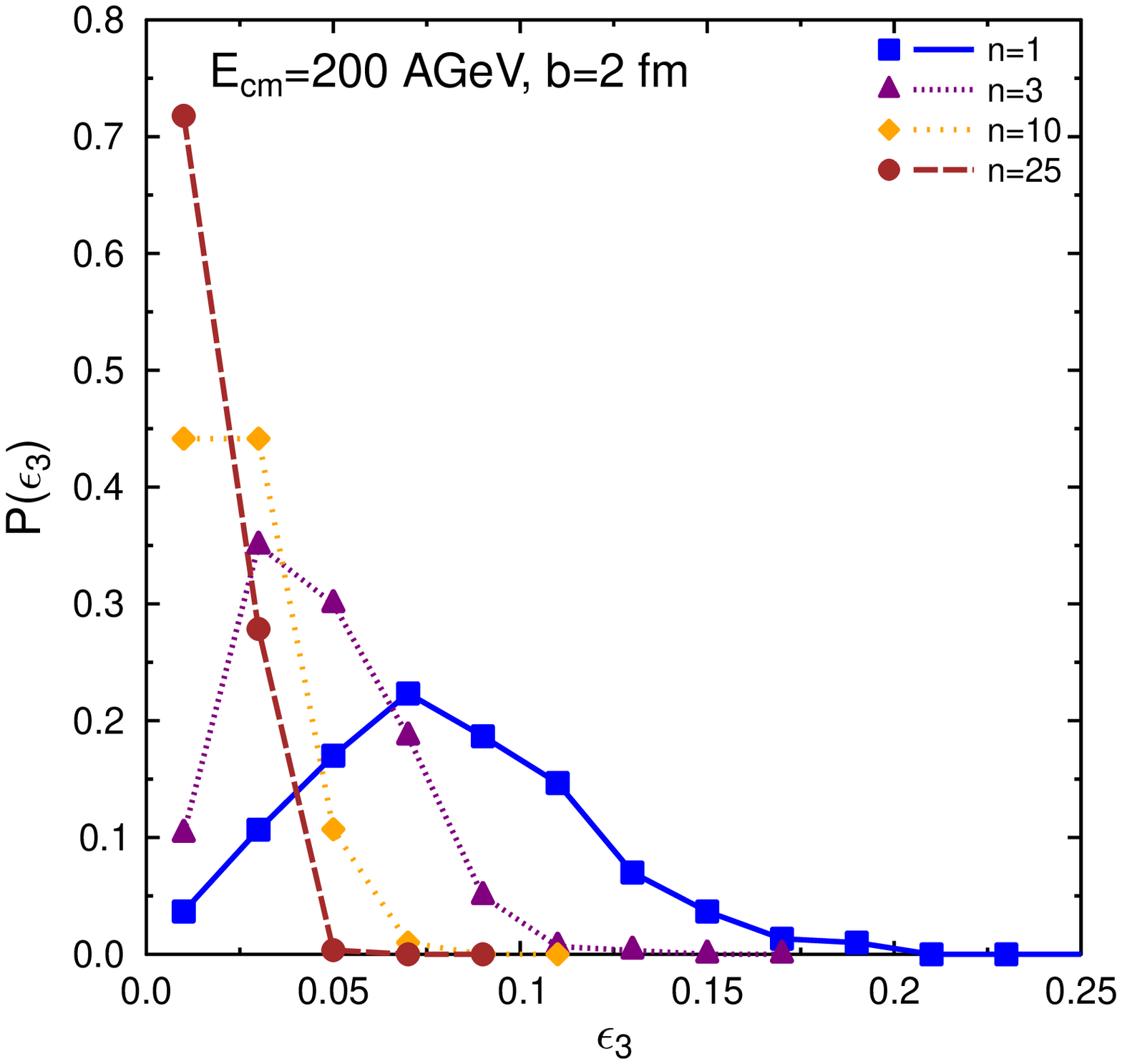}
\includegraphics[width=0.5\textwidth]{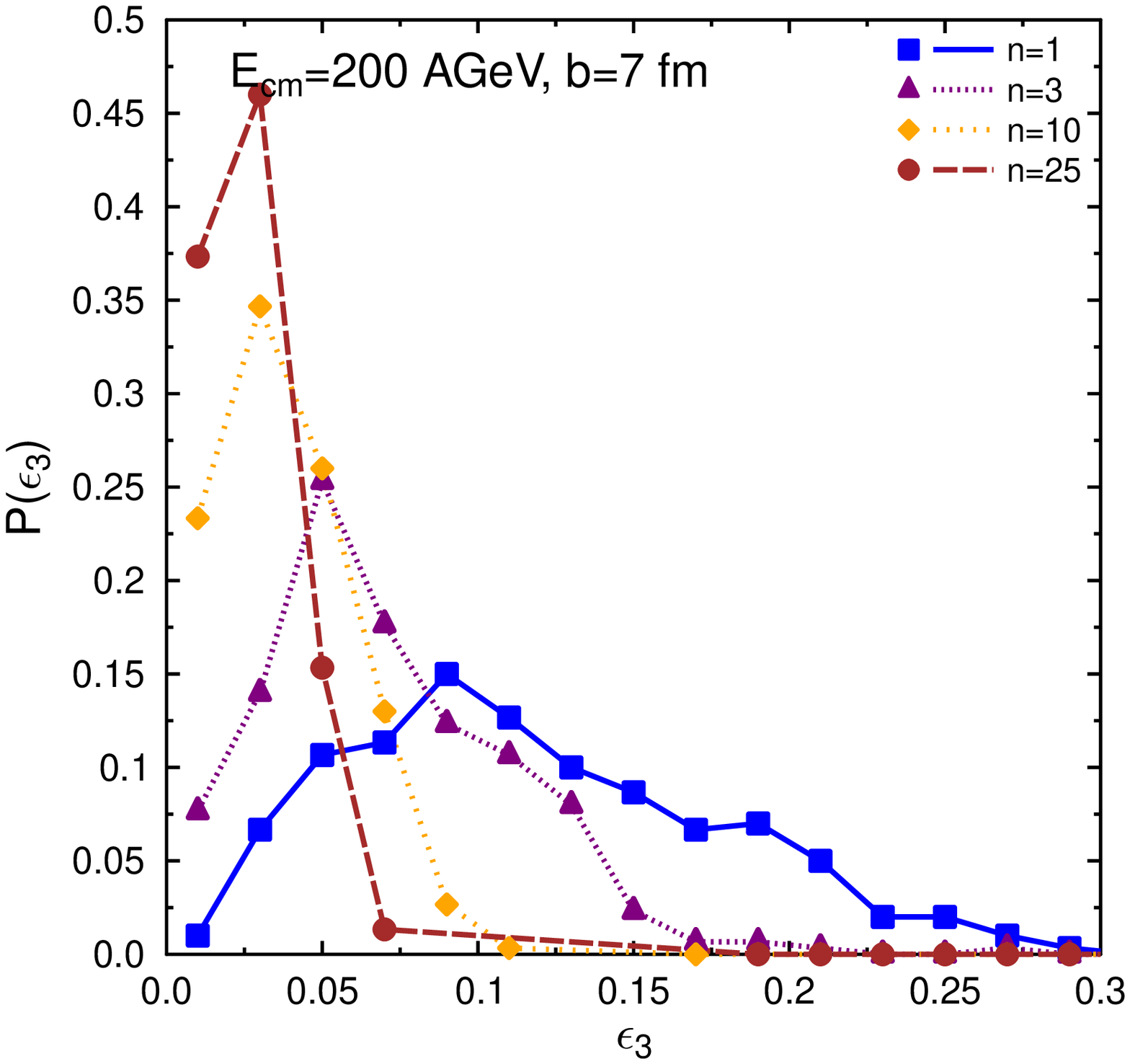}
\caption[Triangularity Distributions]{Triangularity distributions for different
granularities central/mid-central (left: $b=2$ fm and right: $b=7$ fm) Au+Au collisions at
$\sqrt{s_{\rm NN}}=200$ GeV.}
\label{fig_tri_dist} 
\end{figure}

Triangular flow and the initial state triangularity are only generated by initial state fluctuations independent of centrality (as long as the collision is not ultra-peripheral \cite{Nagle:2010zk}). Therefore, the mean values and the general behavior of the triangularity distributions shown in Fig. \ref{fig_tri_dist} are only slightly dependent on centrality. The triangularity decreases to almost zero for smooth initial conditions and shows a clear sensitivity to the granularity. These observations can be summarized in the following statement: The mean value of the triangularity and the fluctuations of the eccentricity are sensitive to the initial state granularity. Since fluctuations of elliptic flow are harder to measure and to compare consistently to model calculations, it seems more promising to look at triangular flow to constrain the amount of fluctuations in the initial state.  

\section[Anisotropic Flow]{Anisotropic Flow Results}
\label{vn_data}
\begin{figure}[h]
\includegraphics[width=0.5\textwidth]{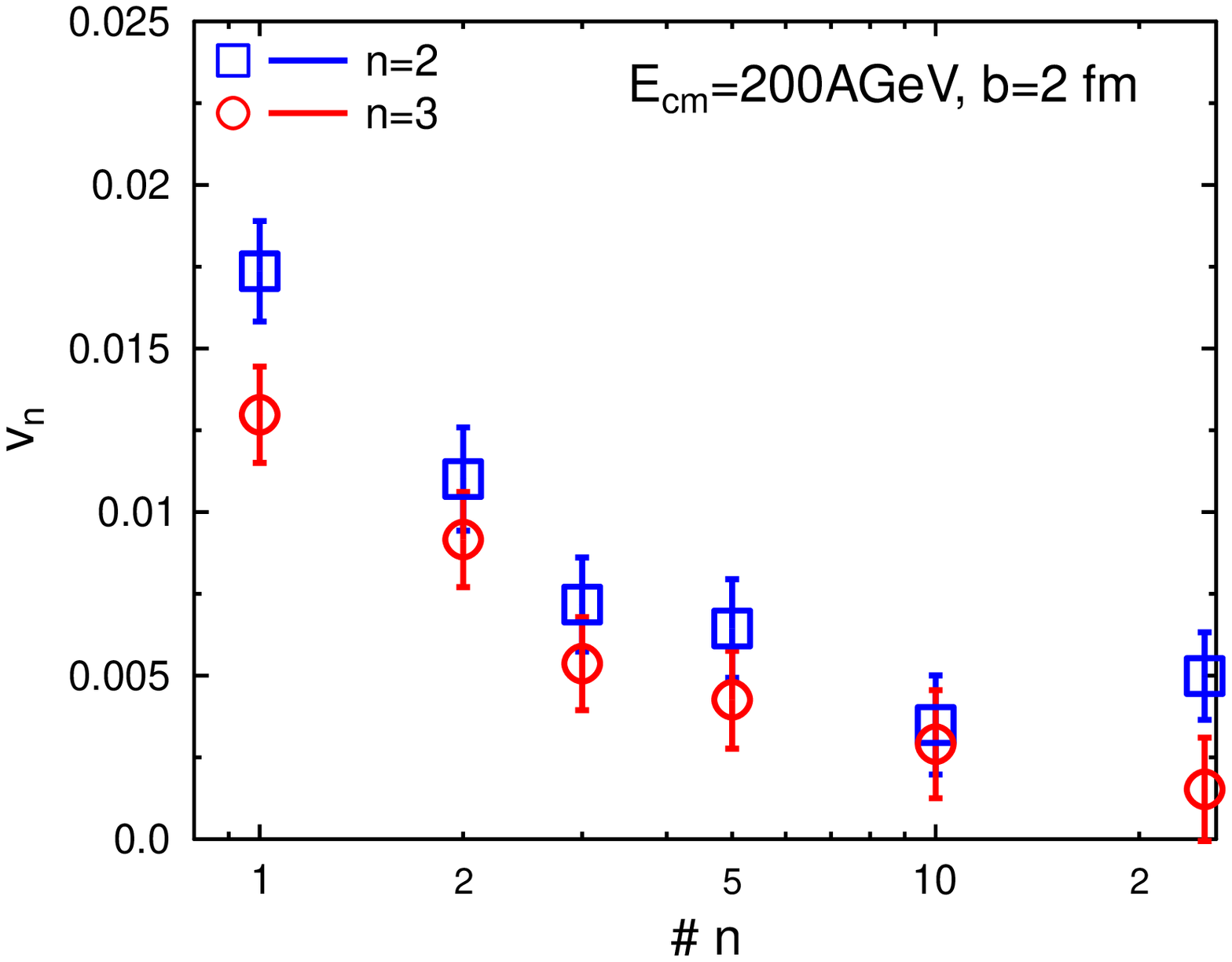}
\includegraphics[width=0.5\textwidth]{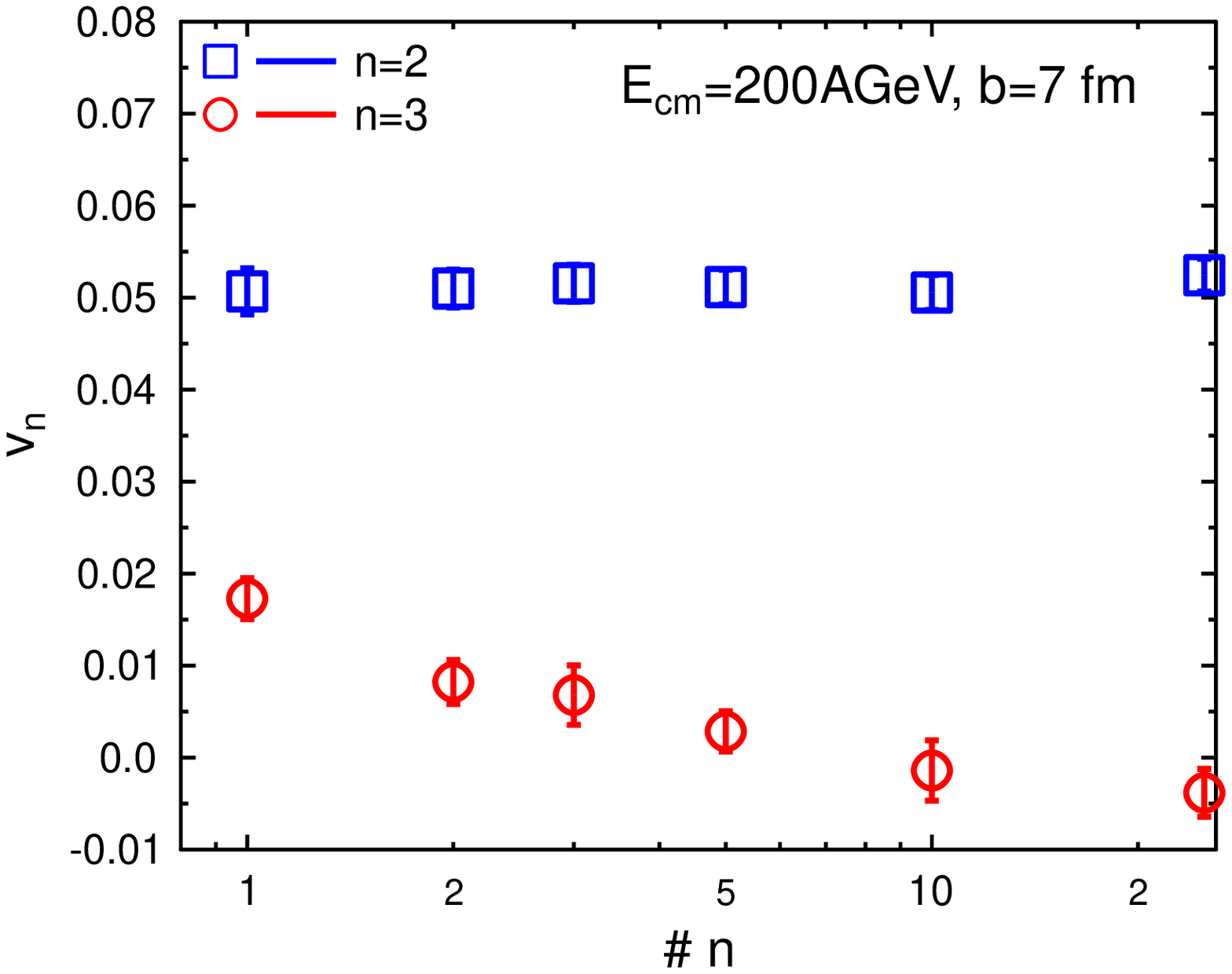}
\caption[Anisotropic Flow]{Anisotropic flow for different granularities in
central/mid-central (left: $b=2$ fm and right: $b=7$ fm) Au+Au collisions at
$\sqrt{s_{\rm NN}}=200$ GeV.}
\label{fig_vn_n} 
\end{figure}

In this Section we present results for the averaged $v_{2,3}$ coefficients for charged particles calculated for different granularities ($n=1,...,25$) via the event plane method. The results in Fig. \ref{fig_vn_n} have not been corrected by the resolution factor since the resolution gets very small for smooth initial conditions (with the exception of $v_2$ in non-central collisions), where the flow coefficients almost vanish anyhow.

The clear dependency of the mean values of triangular flow and elliptic flow in central collisions on the number of events over which the initial average has been performed proves that one can constrain the granularity of the initial state by comparing flow coefficients of calculations to experimental data. 

\begin{figure}[h]
\includegraphics[width=0.5\textwidth]{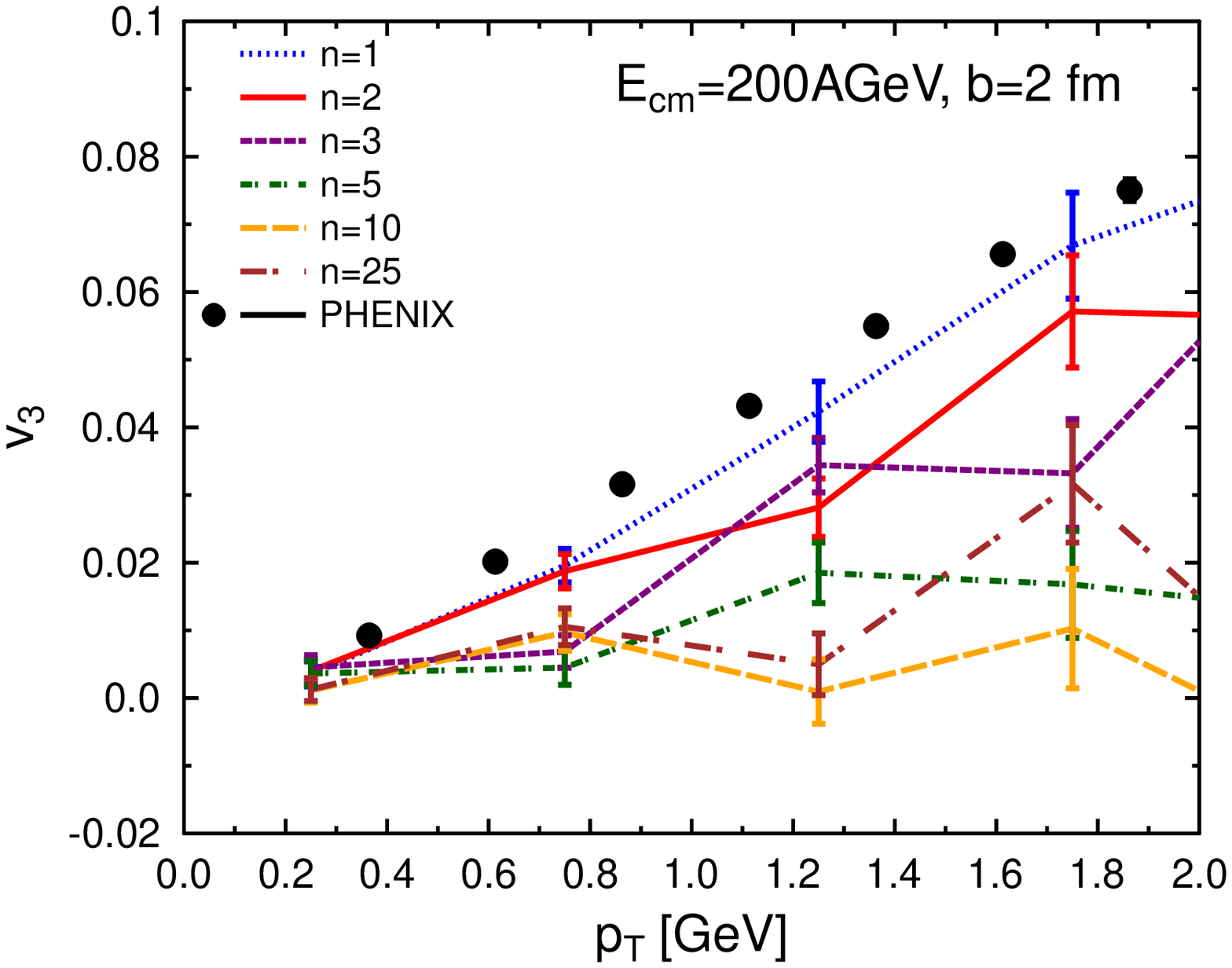}
\includegraphics[width=0.5\textwidth]{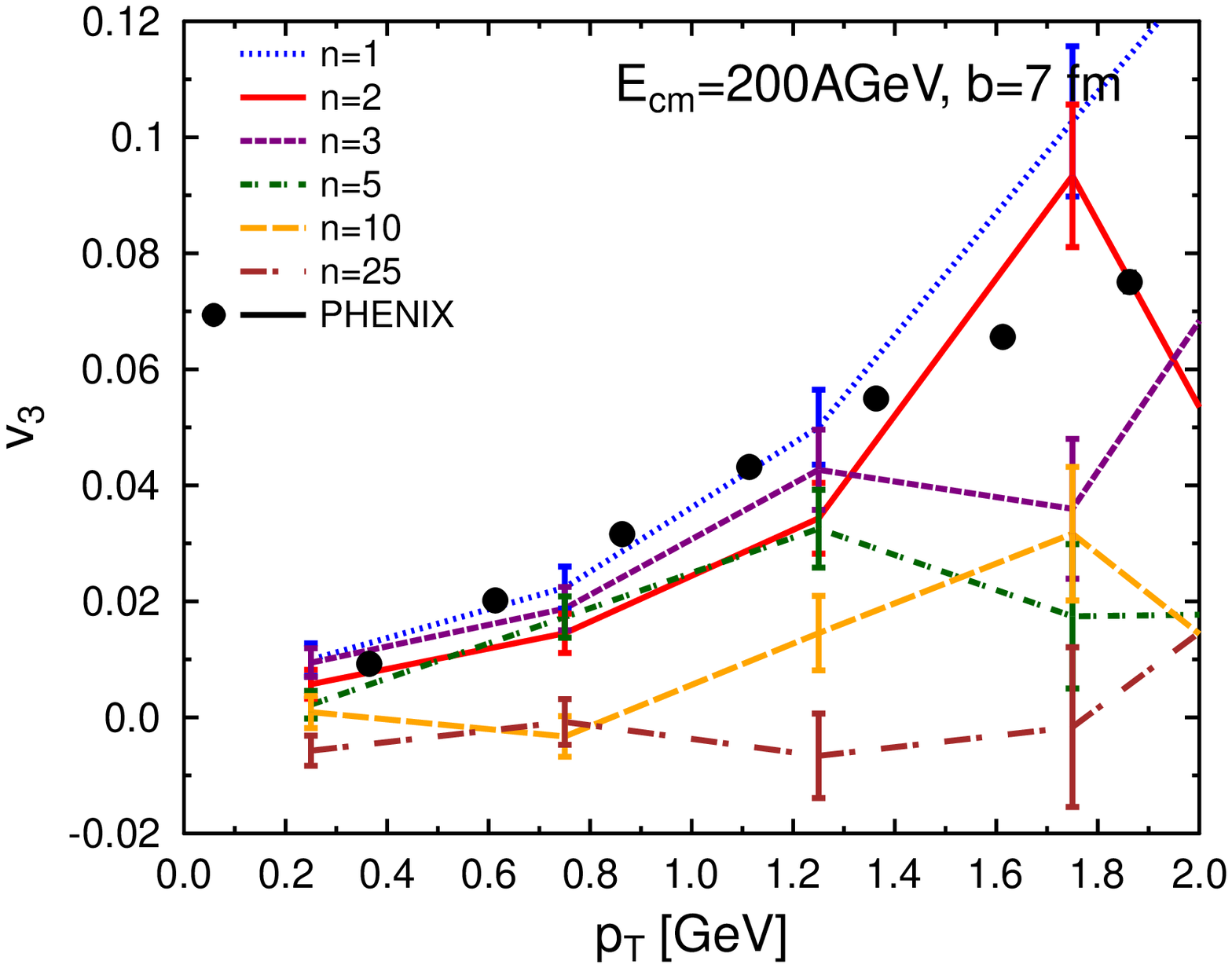}
\caption[Triangular Flow]{Triangular flow versus transverse momentum for
different granularities central/mid-central (left: $b=2$ fm and right: $b=7$ fm) Au+Au collisions at
$\sqrt{s_{\rm NN}}=200$ GeV compared to PHENIX
data \cite{Adare:2011tg}.}
\label{fig_v3_pt} 
\end{figure}

To get a first impression on the amount of fluctuations that is consistent with the PHENIX data on triangular flow \cite{Adare:2011tg}, the transverse momentum dependence of triangular flow of charged particles has been calculated in the hybrid approach (see Fig. \ref{fig_v3_pt}). Up to a value of $n=5$ the resolution of the event plane has reasonable values and the calculation has been corrected by the resolution factor. For the two smoother cases again the resolution gets too small and the triangular flow is practically zero. This first comparison to data indicates that the full event-by-event setup is close to the amount of initial state fluctuations that is necessary. 

Since additional viscosity during the hydrodynamic evolution would dilute fluctuations faster this is a lower bound on the initial state granularity. If there would be a similar event-by-event approach available with a viscous hydrodynamic evolution one could first fit the viscosity to elliptic flow in non-central collisions and then tune the initial state granularity to match the triangular flow data.   

\section[conclusions]{Conclusions and Outlook}
\label{concl}

In this systematic study of we have shown that the mean value of triangular flow can be used to constrain the initial state granularity in heavy ion reactions. A (3+1) dimensional hybrid transport approach that is based on UrQMD including an ideal hydrodynamic expansion has been employed. The granularity of the initial state can be varied by averaging over different numbers of events before generating the energy density distribution for the hydrodynamic evolution. This setup has the main advantage that the bulk properties of the system are not affected and match the experimental data. It is shown that the full event-by-event setup gives currently the best description of the available PHENIX data. 

One future direction of research is to define a measure of the initial state granularity with a more intuitive interpretation, that can be applied to arbitrary initial state profiles. The wealth of experimental data on higher order flow coefficients in turn will help to pin down this granularity measure and define the initial state profiles. Our plan is to perform a multi-paramater sensitivity analysis using sophisticated statistical methods for that purpose \cite{inprep}. 

\section*{Acknowledgements}
We are grateful to the Open Science Grid for the computing
resources. The authors thank John Chin-Hao Chen for providing the data tables of the experimental data and Dirk Rischke for
providing the 1 fluid hydrodynamics code. H.P. acknowledges a Feodor Lynen
fellowship of the Alexander von Humboldt
foundation. This work was supported in part by U.S. department of Energy grant
DE-FG02-05ER41367. Rolando La Placa was supported by the TUNL-REU program via the 
U.S. National Science Foundation (Grant No. NSF-PHY-08-51813).

\end{document}